\documentclass[useAMS,usenatbib,usegraphicx]{mn2e}

\usepackage{amsmath}

\title[Super-solar metallicity in a massive $z\sim1.4$ early-type galaxy]{Old age and super-solar metallicity in a massive $z\sim1.4$ early-type galaxy from
VLT/X-Shooter spectroscopy\thanks{Based on observations collected at the European Organisation for Astronomical Research in the Southern Hemisphere, Chile (program:
086.A-0088(A))}}
\author[I. Lonoce et al.]{I. Lonoce$^{1,2}$\thanks{E-mail: ilaria.lonoce@brera.inaf.it (OAB)}, M. Longhetti$^{1}$, C. Maraston$^{3}$, D. Thomas$^{3}$, C. Mancini$^{4}$, 
A. Cimatti$^{5}$,
\newauthor F. Ciocca$^{1,2}$, A. Citro$^{5}$, E. Daddi$^{6}$, S. di Serego Alighieri$^{7}$, A. Gargiulo$^{5}$, R. Maiolino$^{8}$,
\newauthor F. Mannucci$^{7}$, M. Moresco$^{5}$, L. Pozzetti$^{5}$, S. Quai$^{5}$, P. Saracco$^{1}$\\
$^{1}$INAF-Osservatorio Astronomico di Brera, via Brera 28, I-20121 Milano, Italy\\
$^{2}$Dipartimento di Scienza e Alta Tecnologia, Universit\`a degli Studi dell'Insubria, via Valleggio 11, I-22100 Como, Italy\\
$^{3}$Institute of Cosmology and Gravitation, University of Portsmouth, Dennis Sciama Building, Burnaby Road, Portsmouth PO1 3FX, UK\\
$^{4}$INAF - Osservatorio Astronomico di Padova, Vicolo dell'Osservatorio 5, I-35122 Padova, Italy \\
$^{5}$Dipartimento di Fisica e Astronomia, Universit\`a di Bologna, Viale Berti Pichat 6/2, I-30127 Bologna, Italy \\
$^{6}$CEA-Saclay, Service d'Astrophysique, F-91191 Gif-sur-Yvette, France\\
$^{7}$INAF---Osservatorio Astrofisico di Arcetri, Largo Enrico Fermi 5, I-50125 Firenze, Italy \\
$^{8}$Kavli Institute for Cosmology, University of Cambridge, Madingley Road, Cambridge CB3 0HA, UK }
\begin{document}

\date{Accepted . Received ; in original form }

\pagerange{\pageref{firstpage}--\pageref{lastpage}} \pubyear{2015}

\maketitle

\label{firstpage}

\begin{abstract}
\noindent
We present the first estimate of age, \emph{stellar} metallicity and chemical abundance ratios, for an individual early-type galaxy at high-redshift (\emph{z} $=1.426$) in the COSMOS field. 
Our analysis is based on observations obtained with the X-Shooter instrument at the VLT, which cover the visual and near infrared 
spectrum at high (R $>5000$) spectral resolution. We measure the values of several spectral absorptions tracing chemical species, in particular Magnesium and Iron, besides determining the
age-sensitive D$4000$ break. 
We compare the measured indices to stellar population models, finding good agreement. 
We find that our target is an old ($t>3$ Gyr), high-metallicity ([Z/H] $>0.5$) galaxy which formed its stars at \emph{z$_{form}$} $>5$ within a short time scale $\sim0.1$ Gyr, as 
testified by the strong [$\alpha$/Fe] ratio ($>0.4$), and has passively evolved in the first $>3-4$ Gyr of its life. 
We have verified that this result is robust against the choice and number of fitted spectral features, and stellar population model. 
The result of an old age and high-metallicity has important implications for galaxy formation and evolution confirming an early and rapid formation of the most massive galaxies in the Universe.

\end{abstract}
\begin{keywords}
galaxies: elliptical and lenticular, cD 
-- galaxies: high-redshift -- galaxies: stellar content.
\end{keywords}

\section{Introduction}

The cosmic history of galaxy mass assembly represents one of the open key questions in cosmology. Early-type galaxies (ETGs) are the most effective probes to investigate this
topic, as they are the most massive and oldest galaxies in the local Universe and most likely those whose stars formed earliest.
Observations have shown that a population of massive and passive galaxies is already in place at high redshift, when the Universe was only a few Gyr old (\citealt{cimatti04},
\citealt{saracco05}). 
So far, the main physical parameters related to their formation and assembly have been mainly estimated on local ETGs, and their ageing and evolution can mix up and 
confuse the original properties when the bulk of their mass formed and assembled.
Information on the star formation (SF) time-scale of high-\emph{z} ETGs can be obtained from the detailed chemical  abundance ratios of their stellar
populations \citep{thomas05}, which can be derived by a detailed spectral analysis. 
Indeed, the abundance of Iron with respect to $\alpha$-elements is tightly correlated with the time delay between Type I and Type II supernovae (SN), giving a
direct probe of the time-scale within which SF has occurred.

Up to now, only few works have experimented a spectral analysis on ETGs at \emph{z} $>1$
(\citealt{mio}, \citealt{onodera}, \citealt{jorgensen}) due to the low S/N of the available spectroscopic data, and they were mostly focused on age estimates, in particular using the $UV$ region \citep{cimatti08}. 
The analysis of the rest-frame optical spectrum is still lacking. 

Furthermore, measures are usually performed on stacked spectra \citep{onodera15}, thus deleting possible peculiarities of single objects. 

A single-object measurement of age, stellar metallicity and chemical abundance ratios of \emph{z} $>1.2$ ETGs is missing at the present time. We fill this gap, presenting the first attempt to measure the detailed chemical composition, besides age, of a \emph{z} $\sim1.4$ ETG directly in the early stages of its evolution.

Throughout this paper, we assume a standard cosmology with H$_0=70$ km s$^{-1}$ Mpc$^{-1}$, $\Omega_m=0.3$ and $\Omega_{\Lambda}=0.7$.

\section{COSMOS-307881: spectroscopic data}
\label{object}

\begin{figure*}
\includegraphics[width=16.7cm]{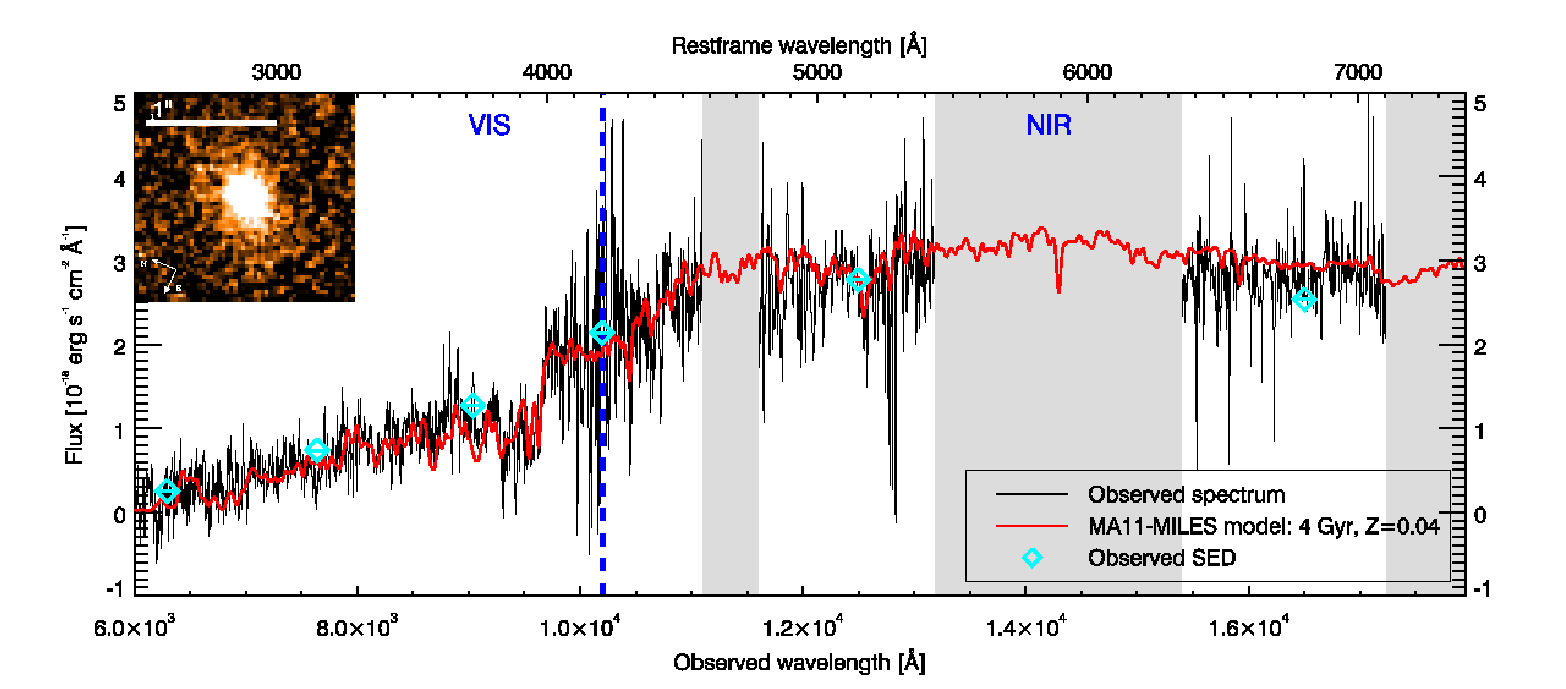}
\includegraphics[width=16.7cm]{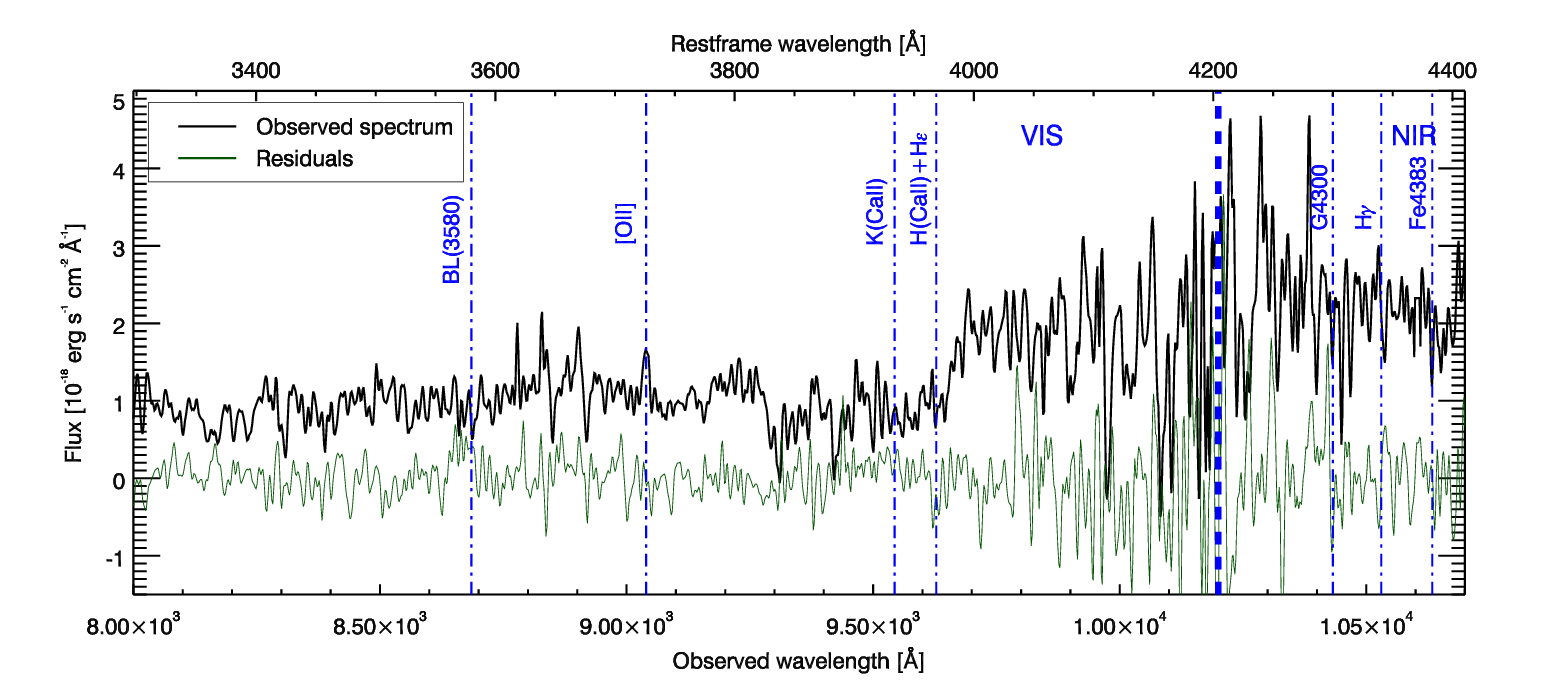}
\includegraphics[width=16.7cm]{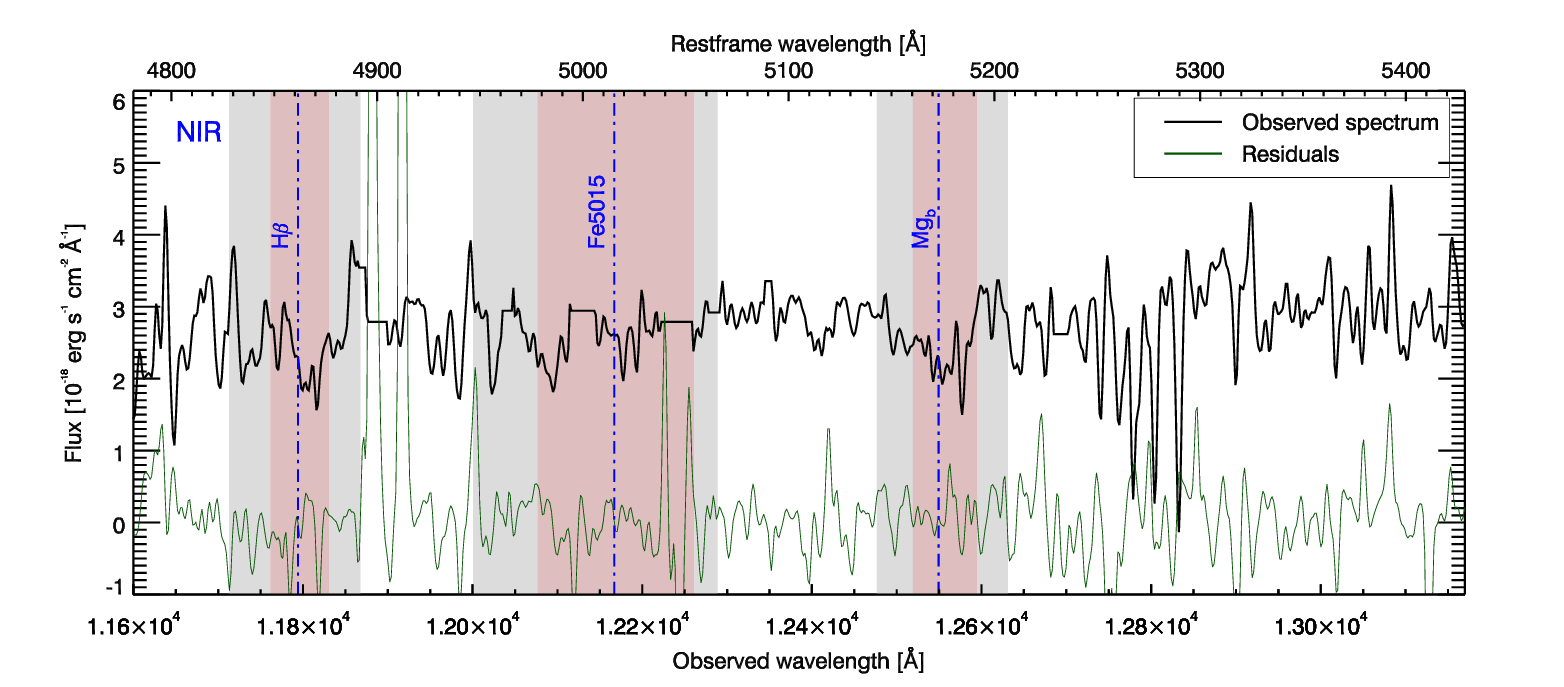}
\caption{COSMOS $307881$. The galaxy observed spectrum (black line) is compared to a model \citep{ma11} with age of $4$ Gyr and super-solar metallicity (Z$=0.04$), and to observed photometric data (cyan diamond). Top panel: VIS and NIR spectral region together with the HST/ACS I-band image of the
target. Middle panel: zoom of the $4000$ \AA $ $ rest-frame region. Bottom panel: zoom of the $5000$ \AA $ $ restframe region. The main absorption (and one emission) lines in each spectral 
region are highlighted. Dark green lines indicate the residual spectrum.}
\label{spettro}
\end{figure*}

Our target is a bright and massive ($>10^{11}$ M$_{\odot}$) ETG from the K-selected galaxy catalog in the COSMOS (Cosmological Evolution Survey) field
\citep{cracken}. It is one of the $12$ galaxies with K$_s$(Vega) $<17.7$ selected by \citet{mancini} on the basis of three criteria: (i) non-detection at $24$ $\mu$m in the 
\emph{Spitzer}+MIPS data \citep{sanders}; (ii) visual elliptical morphology (see Fig. \ref{spettro}, top panel); (iii) multicolor Spectral Energy Distribution (SED) consistent 
with old and passive stellar populations with no dust reddening. All available information are shown in Table \ref{tab:info}. For further details see 
\citet{mancini} and \citet{onodera}.
\begin{table*}
 \begin{minipage}{165mm}
 \caption{COSMOS-307881. Data derived from the analysis of \citet{onodera}: K-band magnitude in Vega system (K$_s$); spectroscopic redshift ($z_{spec}^{ONOD}$); 
 stellar population age (Age$_{phot}$) and logarithm of the stellar mass  (log$\mathcal{M}_*$) derived from SED fitting assuming a Chabrier IMF \citep{chabrier}; 
 effective radius (R$_e$); degree of compactness (C=R$_e$/R$_{e,z=0}$); Sersic index (\emph{n}). Units of right  ascension are hour, minutes and seconds, and unites of declination 
 are degrees, arcminutes and arcseconds.}
 \label{tab:info}
 \begin{tabular}{lccccccccc}
 \hline
 ID & RA & DEC & K$_s$       & \emph{z}$_{spec}^{ONOD}$(*) & Age$_{phot}$ &  log$\mathcal{M}_*$   & R$_e$   & C  & Sersic \emph{n} \\
    &    &     &    (Vega)   &                             & (Gyr)        &   (M$_{\odot}$)       & (kpc)   &	 &   \\
 \hline
 307881 & 10:02:35.64 & 02:09:14.36 & 17.59 & 1.4290$\pm$0.0009 & 3.50  &   11.50 & 2.68$\pm$0.12 & 0.32 &  2.29$\pm$0.10\\
 \hline
\end{tabular}
 (*)This work: \emph{z}$_{spec}=1.426\pm0.001$\\
\end{minipage}
\end{table*}
From these studies we can infer that our target is an old and slightly compact (R$_e\sim0.3$ R$_e^{z=0}$, w.r.t. the local size-mass relation) ETG, and most of its mass is 
composed by a passively evolving stellar population.
In the following we present a new analysis of the stellar population properties of this ETG based on new spectroscopic data, which, thanks to the high resolution in a wide spectral 
range of the X-Shooter spectrograph, allows us to measure the stellar metallicity of this high-$z$ target.

Spectroscopic observations were carried out with the X-Shooter spectrograph on the VLT/UT2 \citep{vernet} in Italian guaranteed time during the nights $9-10$ February 2011, (program: 086.A-0088(A)).
The target has been observed under bad sky condition (seeing $>2''$) during the first night (about $1.7$h for the VIS and UVB arms and about $1.9$h for the NIR arm), while during the second night more
than $4$ hours of exposure have been collected under good sky conditions (seeing $\sim0.8''$). We decided to consider only the latter set of observations. The use of the
$1''.0$ slit in the UVB arm and $0''.9$ slit in the  VIS and NIR arms, resulted in a spectral resolution of $5100$, $8800$ and $5600$ in the UV, VIS and NIR respectively.

The data reduction has been performed taking advantage of the ESO pipeline \citep{goldoni} regarding the first steps of the process, and completed by means of {\small\texttt{IRAF}} tools. 
In Fig. \ref{spettro} the reduced mono-dimensional spectrum of $307881$ (black line) is shown in comparison with the photometric points (cyan diamonds, from \citealt{muzzin}, 
and \citealt{mccracken}). Overlaid is a 4 Gyr, super-solar metallicity (Z$=0.04$) simple stellar population model of \citet{ma11} (M11) based on the MILES stellar library \citep{miles} (red line).
Note that this template is the most similar in term of age and total metallicity, to the best-fit solution that will be derived in the next section. The full spectral model, as known, does
not include the $\alpha$/Fe parameter (see \citet{ma11}), which we shall derive using selected absorption-line models in the next section (see \citet{tmj}).

The original high resolution of the spectrum (e.g. $\sim1$ \AA $ $ in the VIS) has been decreased to match the lower resolution of the models (i.e. $\sim2.54$ \AA $ $ at rest-frame, and $\sim6$ \AA $ $  in the VIS
at \emph{z}$=1.4$), which are used to perform the analysis 
of spectral indices (for details see Lonoce et al. $2015$ in preparation). Note that we adopted the MILES-based M11 models instead of the higher resolution ones (M11-ELODIE or M11-MARCS, see www.maraston.eu/M11) to take advantage of the
increasing S/N in downgrading the observed resolution.
The value of the signal to noise ratio (S/N) obtained in the spectral region around 
$5000$ \AA $ $ restframe is $\sim7$ per pixel.

\begin{figure*}
\includegraphics[width=5.5cm]{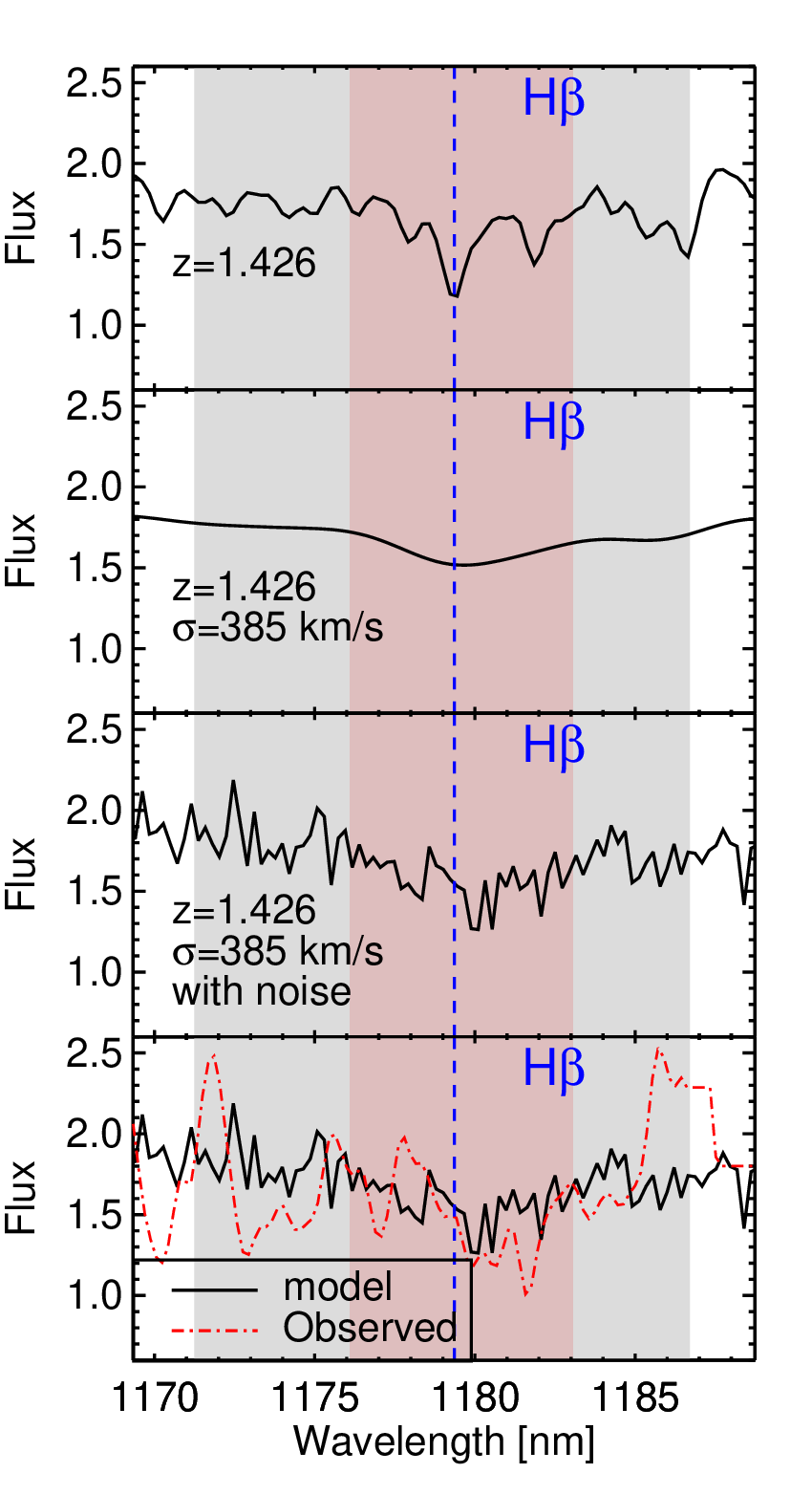}
\includegraphics[width=5.5cm]{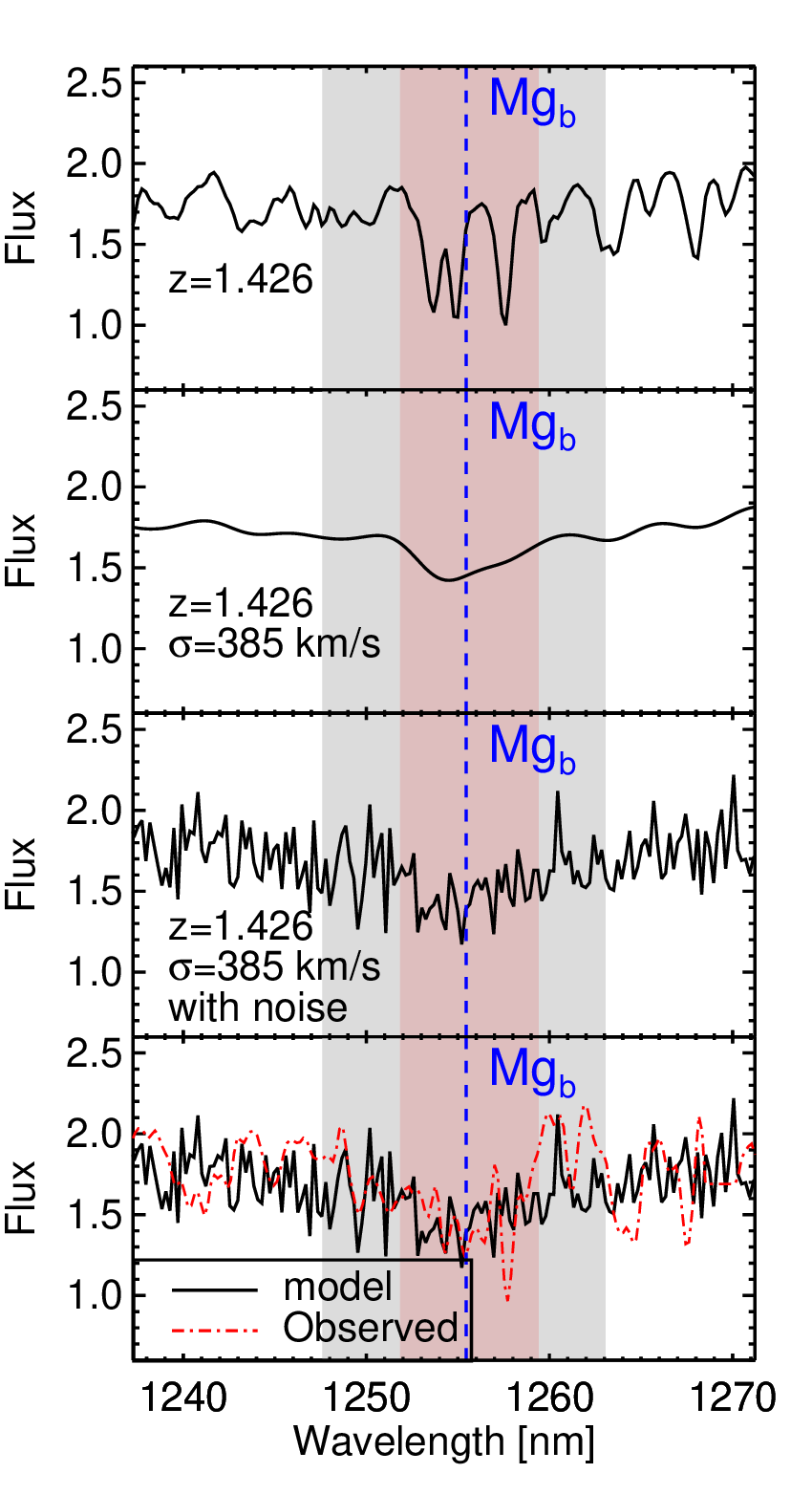}
\caption{H$\beta$ (left panels) and Mg$_b$ (right panels) features in the $4$ Gyr, $2$Z$\odot$ metallicity model shown in Figure 1. From top to bottom we show: model at \emph{z} $\sim1.4$; model corrected for $\sigma=385$ km/s; model downgraded for the observed poissonian noise; model compared to the observed spectrum
(point-dashed red line).}
\label{MgbHb}
\end{figure*}

In Fig. \ref{MgbHb} we show how the H$\beta$ and Mg$_b$ are expected to appear on a synthetic template (upper first panel) that
broadly reproduces the observed one. The model is firstly adapted to the measured velocity dispersion (upper second panel), then downgraded to be as noisy as the observed spectrum (upper third
panel) and finally compared to the observed one (bottom panel). The similarity between expected and observed absorption features is quite evident.

\section{SPECTRAL ANALYSIS}
\label{analisi}
As a first step, we measured the redshift of $307881$ fitting the Mg$_b$ line region, which is the cleanest from the background residuals, as it can be seen in Fig. \ref{spettro} (bottom
panel), finding \emph{z}$=1.426\pm0.001$. 
The Mg$_b$ line region has been also used to find a best fitting velocity dispersion estimate, that resulted to be $\sigma=385\pm85$ km/s. More details on the fitting procedure adopted to 
fix both \emph{z} and $\sigma$ will be described in the forthcoming paper.

We then selected some indices whose absorption features are clearly visible in the observed spectrum, to try to simultaneously derive both the mean age and the metallicity of its 
stellar population. The selected indices are: D$4000$ index \citep{hamilton}, H$\gamma_F$ \citep{wortheyott}, G$4300$, Fe$4383$, Ca$4455$, Fe$4531$, H$\beta$, Fe$5015$ and
Mg$_b$ (Lick/IDS system, \citealt{worthey}).
In particular, it is well known that the Mg$_b$ index is the best \emph{metallicity} and \emph{chemical abundance} dependent index in the region around $5000$ \AA $ $ restframe \citep{korn}. 
In Table \ref{tab:indici} we report the measured values of these indices together with their errors derived by means of \emph{Monte Carlo} simulations set on the uncertainties in the flux
measurements. 

Finally, we want to point out the presence of the [OII]$3727$ emission line (Fig. \ref{spettro}, middle panel). We can exclude that its origin is due to an active AGN, since we do not see
any other signature in the observed wide spectral window. There are reasonable possibilities that this emission is caused by the UV ionizing emission of old
stars in post main-sequence phases \citep{yi}, as confirmed by UV indices (Lonoce et al. $2015$ in preparation), while a strong contribution from SF can be excluded.     
Indeed, as it can be noted from Fig. \ref{MgbHb} (left panels), it is highly unlikely that the H$\beta$ feature is affected by emission, considering also that its value suggests a 
 stellar population age in good agreement with that derived by the H$\gamma$ and  D$4000$ index.

\begin{table}
 \centering
 \caption{Measured indices values.}
 \begin{tabular}{lr}
 \hline
 Index & value \\
 \hline
 D$4000$      & $ 2.44\pm0.12$   \\
 D$_n4000$    & $ 2.42\pm0.17$   \\
 H$\gamma_F$  & $-1.56\pm0.92$  \\
 G$4300$      & $ 6.52\pm1.12$   \\
 Fe$4383$     & $ 7.40\pm1.74$   \\
 Ca$4455$     & $ 1.06\pm0.83$   \\
 Fe$4531$     & $ 3.20\pm1.40$   \\
 H$\beta$     & $ 2.52\pm0.93$  \\
 Fe$5015$     & $ 3.89\pm1.91$   \\
 Mg$_b$       & $ 5.75\pm0.81$   \\
 \hline
\end{tabular}
\label{tab:indici}
\end{table}

\section{Model comparison}
\label{comparison}

\begin{figure}
\begin{centering}
\includegraphics[width=8.5cm]{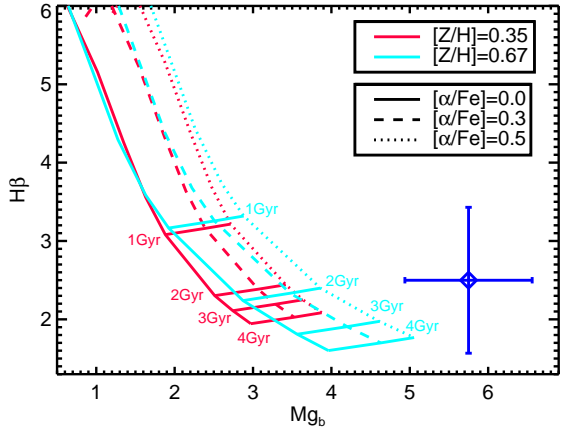}
\caption{H$\beta$ versus Mg$_b$ plot. Comparison between the measured indices (blue diamond) and models of \citet{tmj} (lines). Ages run from $0.1$ to 4 Gyr, for super solar
metallicities of [Z/H]$=0.35$, $0.67$ (red, cyan lines), and $\alpha$-element abundances: [$\alpha$/Fe]$=0.0$, $0.3$, $0.5$ (solid, dashed, dotted lines). The models are
corrected for the measured value of $\sigma=385$ km/s.}
\label{mgb}
\end{centering}
\end{figure}

In Fig. \ref{mgb} we show the observed Mg$_b$ and H$\beta$ indices (blue diamond), compared to the predictions of the SSP of \citet{tmj}, based on the MILES library for a wide range of ages
(from $0.1$ Gyr shown up to the age of the Universe at $z\sim1.4$, i.e. $4$ Gyr); 
super-solar metallicities [Z/H]$=0.35$ (red lines) and [Z/H]$=0.67$ (cyan lines), and various $\alpha$/Fe parameters, namely [$\alpha$/Fe]$=0.0$, $0.3$, $0.5$. 
Models assume a \citet{salpeter} initial mass function
(IMF), and are corrected for the measured velocity dispersion value of $\sigma=385\pm85$ km/s. As it can be seen, the extreme value of the Mg$_b$ index fully requires high metallicity models up to [Z/H]$=0.67$. 
In particular, in Fig. \ref{mgb} the model expectations for these indices are reported also in case of non-solar values of the $\alpha$-enhancement [$\alpha$/Fe], from $0.3$ 
to $0.5$ (dashed and dotted lines respectively). 

Models corresponding to extreme values of [$\alpha$/Fe]$>0.5$ seem to be required to match the observations.

\begin{figure}
\begin{centering}
\includegraphics[width=8.5cm]{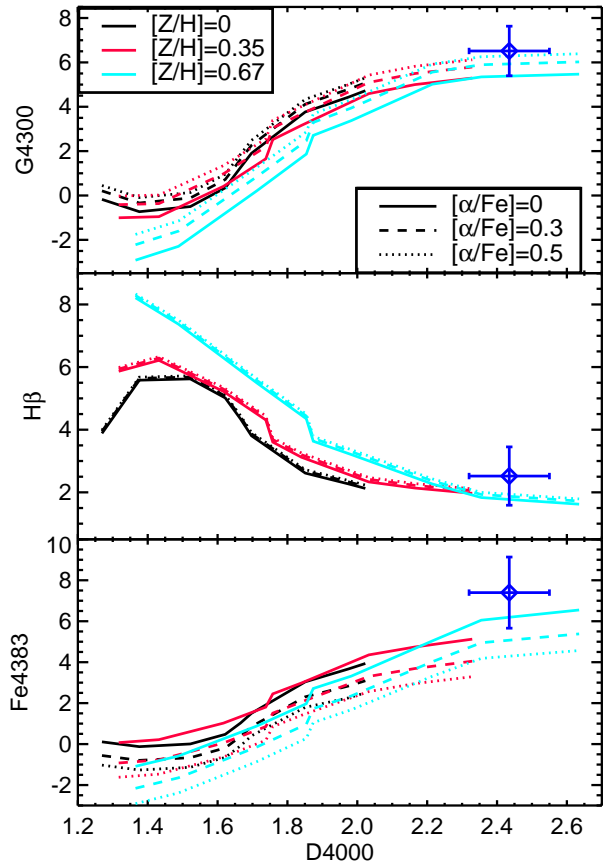}
\caption{Comparison between four measured indices (G4300, H$\beta$ and Fe$4383$ versus D$4000$) and the models of \citet{tmj}. Ages run
from $0.1-4$ Gyr, metallicities from [Z/H]$=0$, $0.35$, $0.67$ (black, red, cyan lines), and $\alpha$-element abundances from [$\alpha$/Fe]$=0$, $0.3$, $0.5$
(solid, dashed, dotted lines). The measured values are shown with a blue diamond. Indices values of models are corrected for the measured value of $\sigma=385$ km/s.}
\label{modelli}
\end{centering}
\end{figure}

The behaviour of the Mg$_b$ index, requiring such extreme values of Z, is confirmed also when
considering the other absorption lines that we were able to measure on this X-Shooter spectrum. Indeed, as it can be noticed in Fig. \ref{modelli} where we propose $3$
examples of Lick indices (G$3400$, H$\beta$ and Fe$4383$, upper, middle and bottom panel respectively) as a function of D$4000$, all measured indices consistently 
point towards a very-high metallicity ([Z/H] $\sim0.67$, cyan lines) being only marginally consistent with the $2$Z$_{\odot}$ values ([Z/H] $\sim0.35$, red 
lines). 

Notice that the highest metallicity models in \citet{tmj} are partly in extrapolation, as they are sampling the edge of the parameter space in terms of the available empirical fitting 
functions for such extreme stellar parameters (see \citet{johan}). At the same time, the underlying stellar tracks are based on real calculations (see
\citet{ma03} for details).

\begin{figure*}
\begin{centering}
\includegraphics[width=12cm]{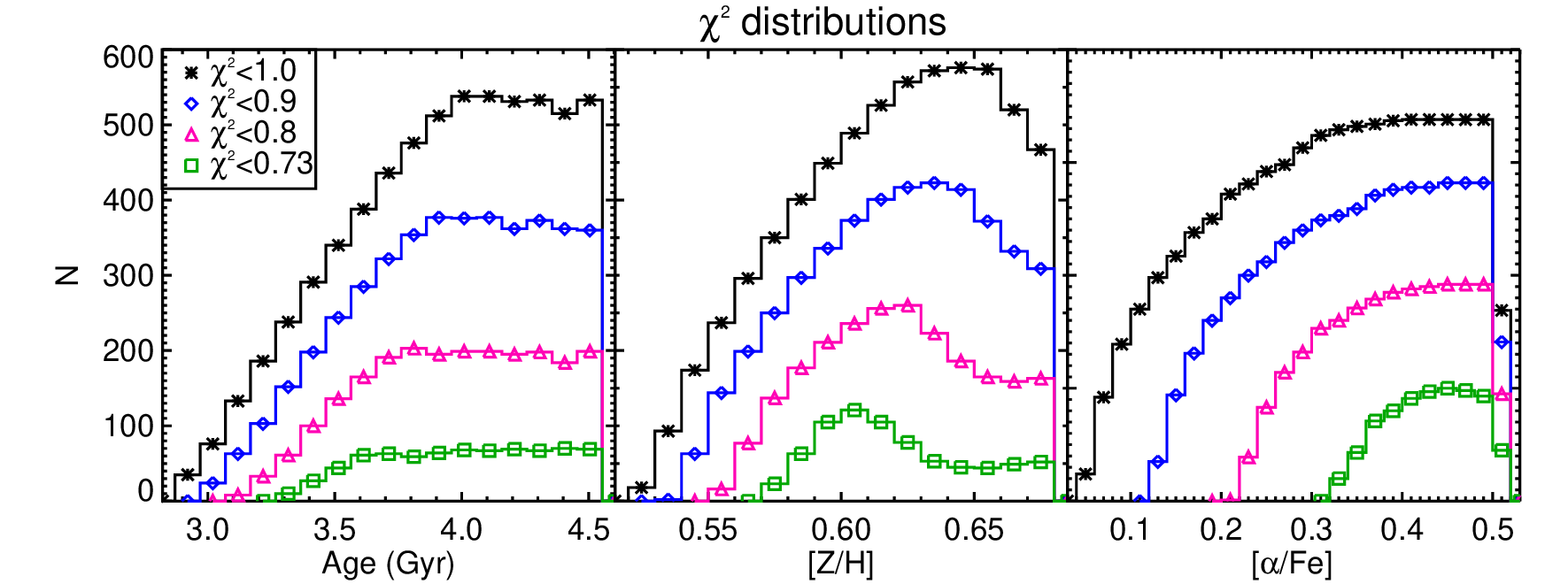}
\includegraphics[width=12cm]{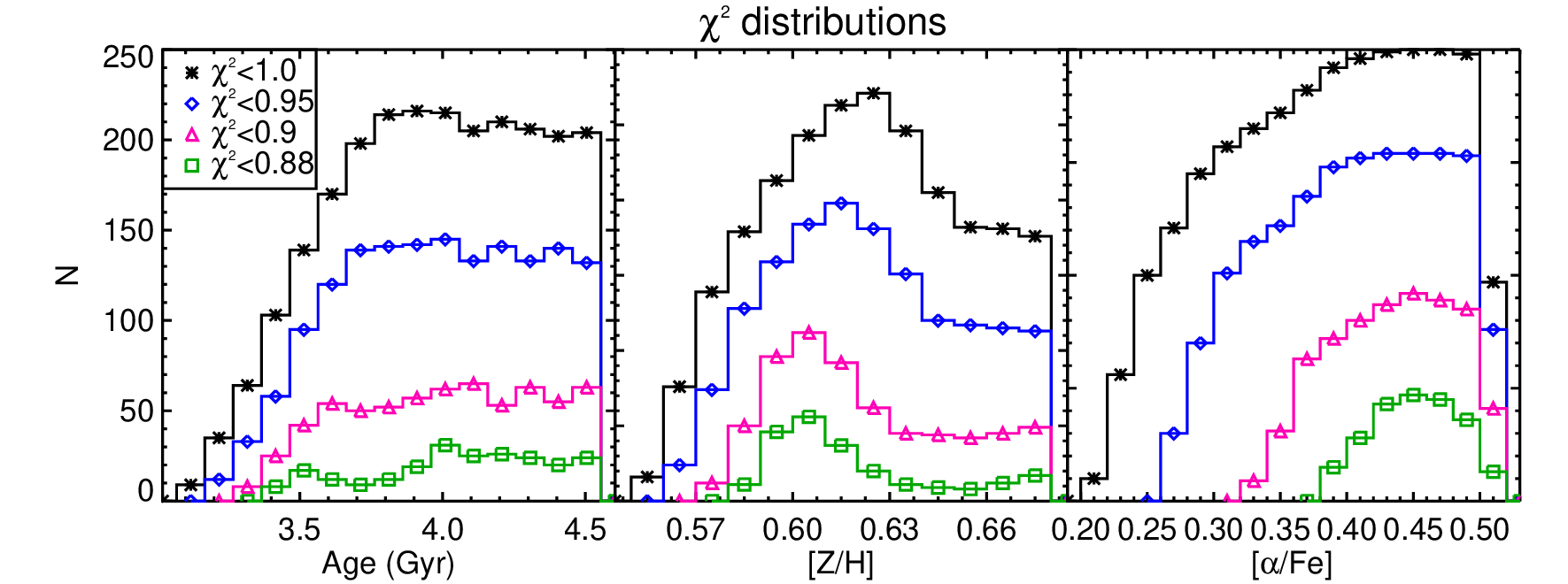}
\includegraphics[width=12cm]{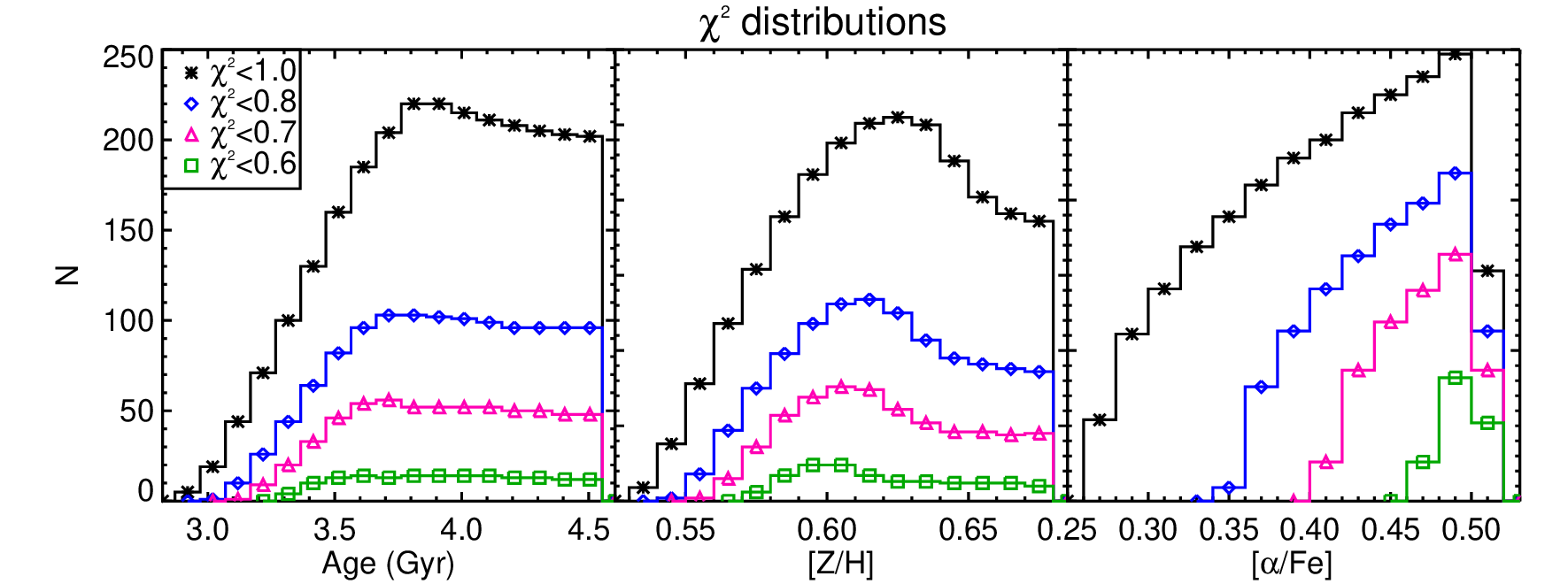}
\caption{Distributions of the fitting-solutions: age, metallicity and $\alpha$-enhancement. Colours and symbols indicate solutions in different $\chi^2$ ranges, e.g. green squares include the most probable
solutions. Top panel: distributions obtained from the all-indices analysis; middle panel: distributions obtained from a smaller set of indices: D$4000$, G$4300$, H$\gamma$, Fe$4383$,
H$\beta$, Fe$5015$ and Mg$_b$; bottom panel: distributions obtained from a smaller set of indices: D$4000$, H$\gamma$, H$\beta$ and Mg$_b$.}
\label{distribuzioni}
\end{centering}
\end{figure*}

\begin{figure*}
\begin{centering}
\includegraphics[width=15cm]{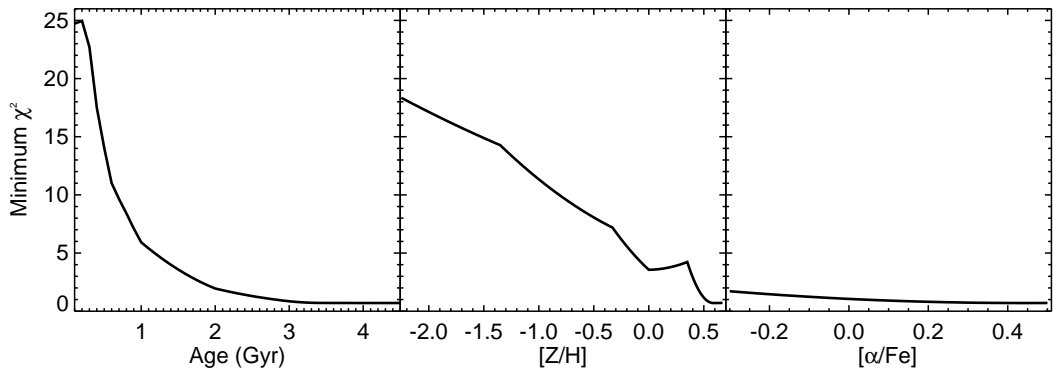}
\caption{Trends of minimum $\chi^2$ for age, [Z/H] and [$\alpha$/Fe] of all obtained solutions. In particular, ages $<2$ Gyr can be completely excluded due to the rapid degrade of $\chi^2$ towards younger ages.}
\label{chi2all}
\end{centering}
\end{figure*}

More quantitatively, we have computed the best fitting solution obtained comparing all $9$ observed indices values (Table \ref{tab:indici}) with models. The free parameters were age 
($0.1-4.5$, truncated at the age of Universe, with step $0.1$ Gyr)\footnote{Note that we have performed the analysis using all available model ages (up to 15 Gyr) thereby ignoring the age 
of the Universe as a constrain. However, as no particular improvement was noticed in the derived quantities, we decided to focus on ages within the age of the Universe at \emph{z} $\sim1.4$}, the total metallicity (from [Z/H]$=-2.25$ to [Z/H]$=0.67$, 
with step $0.01$) and the $\alpha$/Fe-enhancement (from [$\alpha$/Fe]$=-0.3$ to [$\alpha$/Fe]$=0.5$, with step $0.01$). The minimum $\chi^2$ value corresponds to an age of $4.0^{+0.5}_{-0.8}$ Gyr, metallicity [Z/H]$=0.61^{+0.06}_{-0.05}$ and [$\alpha$/Fe]$=0.45^{+0.05}_{-0.19}$, with $\chi^2=0.7$ and an associated
probability $\sim70\%$. Errors indicate the range values of these parameters over all the solutions associated to probabilities larger than $65\%$. The distributions of the $3$ 
fitting parameters, displayed in different $\chi^2$ ranges, are shown in Fig. \ref{distribuzioni}, top panel.
A global picture of the $\chi^2$ values can be seen in Figure \ref{chi2all} where the minimum $\chi^2$ trends of the $3$ parameters of all solutions are shown. 
It is easy to notice that ages $<2$ Gyr can be completely excluded due to the rapid increasing of their $\chi^2$ values toward younger ages. Instead for ages
$>4.5$ (limit of the Universe age, not shown here) the $\chi^2$ values remains practically constant.

Furthermore, we found that models with Z$\le$Z$_{\odot}$ provide a fit of the free parameters with a probability less than $0.1\%$.

We also verified the strength of this result by repeating the same fitting process selecting smaller and different set of indices, finding very similar
solutions with respect to the previous ones based on the whole set of indices. Two examples are shown in Fig. \ref{distribuzioni} (middle and bottom panels): the distributions of the fitting 
solutions are obtained from two smaller sets of indices (\emph{i)} D$4000$, G$4300$, H$\gamma$, Fe$4383$,
H$\beta$, Fe$5015$ and Mg$_b$ and \emph{ii)} D$4000$, H$\gamma$, H$\beta$ and Mg$_b$) which lead to a best-fit solution of \emph{i)} age$=4.0$ Gyr, [Z/H]$=0.61$ and [$\alpha$/Fe]$=0.44$ with
$\chi^2=0.9$, and \emph{ii)} age$=4.0$ Gyr, [Z/H]$=0.60$ and [$\alpha$/Fe]$=0.5$ with $\chi^2=0.5$,
both totally consistent with the all-indices one.

\begin{figure}
\begin{centering}
\includegraphics[width=9.5cm]{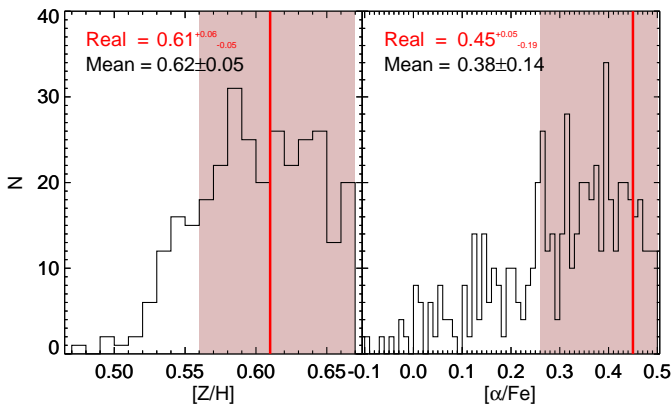}
\caption{Distributions of [Z/H] and [$\alpha$/Fe] derived for a set of $500$ mock spectra built on a model spectrum with parameters (Table 
\ref{tab:results}) as those of the best-fit to the observed spectrum and adding the poissonian noise. The two distributions are well peaked around the true original values (red vertical lines).}
\label{500}
\end{centering}
\end{figure}

We also evaluated the feasibility of this analysis on this low S/N spectrum, in particular for metallicity and $\alpha$-enhancement estimates, by repeating it on a set of $500$ mock spectra built on a model spectrum 
with parameters as the best-fit to the observed one (cfr. Table \ref{tab:results}) and downgraded with the poissonian observed noise. The obtained distributions of the measured [Z/H] and 
[$\alpha$/Fe] are shown in Fig. \ref{500}. They all peak around the true original values (red vertical lines) demonstrating that within the declared errors the obtained values are solid.

Furthermore we tested if the large error on the velocity dispersion could affect the results by performing the same analysis assuming $\sigma=300$ km/s. We found the same best-fit solution
($4.1$ Gyr, [Z/H]$=0.6$, [$\alpha$/Fe]$=0.41$ with $\chi^2=0.6$). 

Finally, in order to test the model dependence of this result, we repeated the same analysis adopting the \citet{bc03} models (BC03). These models do not include the 
[$\alpha$/Fe] parameter, hence we shall use them to constrain age and total metallicity solely. The BC03 models cover a slightly lower $Z$ range with respect to our fiducial 
TMJ models, and are based on different stellar evolutionary tracks. The BC03 best-fit corresponds to an age of $4.5$ Gyr and a metallicity of [Z/H]$=0.4$, which is the maximum 
available metallicity in these models. Hence our result of a high-age and high-$Z$ is not model dependant.

{\it It is important to note the high metallicity is mainly derived as a consequence of the maximum allowed age of $4.5$ Gyr. Should we allow the age to be older than the age 
of the Universe, the metallicity will decrease, as a result of age/metallicity degeneracy. }
Actually, the large error bar of both the indices and of $\sigma$ prevents a real
precise measure of the metallicity, but the peculiarity of such a strong
Mg$_b$ absorption band combined with the narrow range of possible ages (due
to its redshift), make necessary the assumption of a very high value of
the stellar metallicity.


\section{Discussion and conclusions}

\label{discussion}
\begin{table}
 \centering
 \caption{Results: spectroscopic redshift (\emph{z}$_{spec}$) and velocity dispersion ($\sigma$) obtained from this spectroscopic analysis; best-fit values of age (Gyr), [Z/H] 
 and [$\alpha$/Fe].}
 \begin{tabular}{ccccc}
 \hline
 \emph{z}$_{spec}$ & $\sigma$ (km/s) & Age (Gyr) & [Z/H] & [$\alpha$/Fe]  \\
 \hline
 $1.426\pm0.001$ & $385\pm85$ & $4.0^{+0.5}_{-0.8}$ & $0.61^{+0.06}_{-0.05}$ & $0.45^{+0.05}_{-0.19}$ \\
 \hline
\end{tabular}
\label{tab:results}
\end{table}

We have performed a detailed spectroscopic analysis of a \emph{z}=1.426, massive ($M^{*}\sim10^{11} M_{\odot}$), early-type galaxy. We gain strong evidence of a high stellar metallicity and $\alpha$-enhancement, and old (relative to its redshift) age (Table \ref{tab:results}). 
These quantities constrain the past SF history experienced by the galaxy. In particular, the [$\alpha$/Fe] ratio, quantifying the time delay between Type II SN events, responsible of the production of $\alpha$-elements, and 
the Type Ia SNs, related to the formation of Fe-peak elements, allows to determine the SF time-scale \citep{thomas05}. 
The high value [$\alpha$/Fe] $\sim0.4$ obtained for $307881$ is a direct signature that its SF time-scale must have been short. In particular, adopting
the simple theoretical modelling of \citet{thomas05}, where the SF is modelled with a Gaussian function, we obtain a time-scale $\Delta$t $ \sim 0.1$ Gyr covering the interval within which $95\%$ of the stars were formed. Considering also the old age of its stellar content, this suggests that $307881$ formed the bulk of its stars at \emph{z$_{form}$} $>5$ within a short time-scale of $\Delta$t $\sim 0.1$ Gyr and then passively evolved over the following $4$ Gyr. 

With the high [$\alpha$/Fe] value suggesting a short
SF time scale for $307881$, the clear indication for an extremely
high total metallicity opens new issues on the gas enrichment history of the
Universe. 
It is worth emphasising that the global integrated metallicity of
the local ETGs never reaches values higher than the $1-2$ Z$_{\odot}$. Indeed, we
considered the local sample analysed in \citet{thomas10} and quickly verified that the metallicity and 
[$\alpha$/Fe] values of $307881$ together with its velocity dispersion estimate, are not included in the local distribution of values, even if 
such high values of Z and [$\alpha$/Fe] are expected for dense ETGs as suggested by these scaling relations. Thus, this suggests that this galaxy must experience mass 
accretion events (minor merging) from \emph{z} $=1.4$ to
\emph{z} $=0$ which will move it on the observed local scaling relations towards lower values of both Z  
and $\sigma$, diluting the extreme metallicity stars and confining them in the central part of the galaxy. Indeed, 
such extreme metallicity values are in some cases found in the inner core of local massive ETGs (\citealt{thomas05}, \citealt{trager}, \citealt{martin}) which are known to show 
metallicity gradients \citep{labarbera12}. We calculated that adding e.g. $15\%$ of a sub-solar metallicity component, as in the case of dwarf galaxies, 
to $307881$ would decrease the measure of its average metallicity 
to the local observed values. 
On the other hand, the gas metallicities up
to now measured in \emph{z} $>3$ star forming galaxies result to be solar or
sub-solar, and do not match at all the high stellar value measured in our
target galaxy \citep{mannucci} and in the centers of local ETGs \citep{spolaor}.
It is out of the aims of	
the present paper to suggest a possible explanation of this missing
detection of high-\emph{z} high metallicity gas. A possibility is that the lower gas metallicity comes from dilution through infalling primeval gas, or selective mass 
loss of metals in galactic winds. At the same time, the estimate of the metallicity of the Broad Line
Regions in quasars at \emph{z} $>4$ reveals gaseous metallicities even higher than
the stellar one reported here \citep{juarez}, suggesting the possibility
that their enriched gas is involved in the initial SF events
of (at least some) high-\emph{z} massive proto-elliptical galaxies.

\section*{Acknowledgments}

We are grateful to the anonymous referee for helpful comments on our manuscript.
We kindly thank Francesco La Barbera for his precious help in the data reduction and Kyle Westfall for helpful discussions.
IL acknowledges the institute of Cosmology and Gravitation of the University of Portsmouth for a brilliant research visit during which a larger part of this project was worked out. 
IL, ML, AG and PS acknowledge the support from grant Prin-INAF 2012-2013 1.05.09.01.05.
AC, LP, MM acknowledge the support from grant PRIN MIUR 2010-2011. 
CM and DT acknowledge The Science, Technology and Facilities Council for support through the Survey Cosmology and Astrophysics consolidated grant, ST/I001204/1.

\label{lastpage}

\end{document}